\documentstyle[12pt]{article}
\def\figcap{\section*{Figure Captions\markboth
        {FIGURECAPTIONS}{FIGURECAPTIONS}}\list
        {Figure \arabic{enumi}:\hfill}{\settowidth\labelwidth{Figure
999:}
        \leftmargin\labelwidth
        \advance\leftmargin\labelsep\usecounter{enumi}}}
 \relax
\newskip\humongous \humongous=0pt plus 1000pt minus 1000pt
\def\caja{\mathsurround=0pt}
\def\eqalign#1{\,\vcenter{\openup1\jot \caja
        \ialign{\strut \hfil$\displaystyle{##}$&$
        \displaystyle{{}##}$\hfil\crcr#1\crcr}}\,}
\newif\ifdtup

\def\ap#1#2#3{Ann.\ Phys.\ (NY) #1 (19#3) #2}

\def\np#1#2#3{Nucl.\ Phys.\ B#1 (19#3) #2}
\def\pl#1#2#3{Phys.\ Lett.\ #1B (19#3) #2}
\def\pr#1#2#3{Phys.\ Rev.\ D #1 (19#3) #2}
\def\prb#1#2#3{Phys.\ Rev.\ B #1 (19#3) #2}
\def\prep#1#2#3{Phys.\ Rep.\ #1 (19#3) #2}

\def\rmp#1#2#3{Rev.\ Mod.\ Phys.\ #1 (19#3) #2}

\def\cmp#1#2#3{Comm.\ Math.\ Phys.\ #1 (19#3) #2}

\newcounter{hran}
\def\bmini{\setcounter{hran}{\value{equation}}
\refstepcounter{hran} \setcounter{equation}{0}
\renewcommand{\theequation}{\thehran\alph{equation}}
              \begin{eqnarray}  }
\def\bminia{\setcounter{hran}{\value{equation}}
\refstepcounter{hran} \setcounter{equation}{1}
\renewcommand{\theequation}{\thehran\alph{equation}}
              \begin{eqnarray}  }
\def\bminiG#1{
          \setcounter{hran}{\value{equation}}
          \refstepcounter{hran}
          \setcounter{equation}{-1}
          \renewcommand{\theequation}{\thehran\alph{equation}}
          \refstepcounter{equation}
    \label{#1}
          \begin{eqnarray}          }
\def\emini{\end{eqnarray}\setcounter{equation}{\value{hran}}
\renewcommand{\theequation}{\arabic{equation}}}

\newskip\humongous \humongous=0pt plus 1000pt minus 1000pt
\def\caja{\mathsurround=0pt} \def\eqalign#1{\,\vcenter{\openup1\jot
\caja   \ialign{\strut \hfil$\displaystyle{##}$&$
\displaystyle{{}##}$\hfil\crcr#1\crcr}}\,} \newif\ifdtup

\def\half{\mbox{\small $\frac{1}{2}$}}
\def\third{\mbox{\small $\frac{1}{3}$}}
\def\ltap{\raisebox{-.4ex}{\rlap{$\sim$}} \raisebox{.4ex}{$<$}}

\def\frac#1#2{ {{#1} \over {#2} }}

\def\s#1{{\small #1}}

\def\fun#1#2{\lower3.6pt\vbox{\baselineskip0pt\lineskip.9pt
  \ialign{$\mathsurround=0pt#1\hfil##\hfil$\crcr#2\crcr\sim\crcr}}}
\def\ie{\hbox{\it i.e.}{ }}      
\def\eg{\hbox{\it e.g.}{ }}

\def\ds#1{\ooalign{$\hfil/\hfil$\crcr$#1$}}
\def\re#1{(\ref{#1})}

\def\beq{\begin{equation}}
\def\eeq{\end{equation}}
\def\beeq{\begin{eqnarray}}
\def\eeeq{\end{eqnarray}}

\def\eps{\epsilon}
\def\dim{4-n_A-3n_\psi}

\def\s{\sigma}
\def\S{\Sigma}
\def\G{\Gamma}
\def\bG{ \bar \Gamma}

\def\L{ \Lambda}
\def\l{ \lambda}
\def\g{ \gamma}

\def\bx{\mbox{-box}}
\def\bp{ \bar p}
\def\d4#1{\frac {d^4 {#1} }{(2\pi)^4}}

\def\dL{\L{\partial   \over\partial \L}}
\def\DL#1{\L{\partial #1  \over\partial \L}}

\def\dL{\L \partial_\L }
\def\DL#1{\L\partial_\L #1}

\def\UV{$\L_0\to\infty\;$}
\def\IR{$\L\to 0\;\;$}

\def\bit{\begin{itemize}}
\def\eit{\end{itemize}}
\def\ben{\begin{enumerate}}
\def\een{\end{enumerate}}

\def\Maxlp{\raisebox{-1.5 ex}{\rlap{\tiny $\;\;p_i^2<c\l^2$}}
\raisebox{0ex} {$\; \mbox{Max}\;\;\;\,$}}
\def\nome#1{{\label{#1}}}

\begin{document}
\begin{titlepage}
\renewcommand{\thefootnote}{\fnsymbol{footnote}}
\begin{flushright}
     UPRF 93-382 \\
     July 1993
\end{flushright}
\par \vskip 10mm
\begin{center}
{\Large \bf
Ward identities and Wilson renormalization group \\
for QED\footnote{Research supported in part by MURST, Italy}}
\end{center}
\par \vskip 2mm
\begin{center}
        {\bf M.\ Bonini, M.\ D'Attanasio and G.\ Marchesini} \\
        Dipartimento di Fisica, Universit\`a di Parma and\\
        INFN, Gruppo Collegato di Parma, Italy
        \end{center}
\par \vskip 2mm
\begin{center} {\large \bf Abstract} \end{center}
\begin{quote}
We analyze a formulation of QED based on the Wilson renormalization
group. Although the ``effective Lagrangian'' used at any given scale
does not have simple gauge symmetry, we show that the resulting
renormalized Green's functions correctly satisfies Ward identities
to all orders in perturbation theory.
The loop expansion is obtained by solving iteratively the Polchinski's
renormalization group equation.
We also give a new simple proof of perturbative renormalizability.
The subtractions in the Feynman graphs and the corresponding
counterterms are generated in the process of fixing the physical
conditions.
\end{quote}
\end{titlepage}

\section{Introduction}
In a gauge theory the presence of ultraviolet (UV) divergences
could conflict with local gauge symmetry.
There are two cases in which this conflict is avoided:
dimensional regularization, which is applicable only in perturbation
theory, and the Wilson formulation of lattice gauge theories \cite{LGT}
in which the Lagrangian is not local.
However for chiral gauge theories even in these two approaches
one is forced to introduce all possible counterterms,
even those which break the local gauge symmetry \cite{g5,g5LGT}.

The most physical way to deal with UV divergences and to define the theory
even beyond the perturbative expansion is given by the Wilson
renormalization group equations \cite{W}. In this method one
starts with the Lagrangian at the UV scale $\L_0$ which does not
have simple gauge symmetry properties.
Then one deduces the flow of the ``effective Lagrangian'' at
a lower momentum scale $\L$ by performing the path integrals over the
fields with frequencies between $\L$ and $\L_0$.
By integrating over all frequencies (\ie by setting $\L=0$) one obtains
the physical ``effective action'' which should satisfy the Ward or
Slavnov-Taylor identities.
It is interesting to analyze in detail how these identities are violated
along the renormalization group flow but are satisfied for the final
result (\ie $\L=0$ and \UV).

In this paper we study the Wilson renormalization group flow of QED
and analyze how the Ward identities are fulfilled.
We use the formulation given by Polchinski \cite{P} (see also
\cite{G,B} and for recent applications \cite{KK}-\cite{Gir}).
{}From this one deduces a linear differential equation
in the scale $\L$ for a functional $\G[A,\psi,\bar\psi;\L]$
of the QED classical fields $A_\mu$,$\psi$ and $\bar\psi$
(for the case of a scalar theory see Refs.~\cite{BDM,Wet}).
The expansion coefficients of this functional are the ``cutoff
proper vertices'' which are obtained  by performing the path integrals
over the fields with frequencies between $\L$ and $\L_0$, \ie by
cutting off the fields with frequencies outside this range.
One has therefore that the physical effective action, given by performing
the path integrals over all frequency range, are formally obtained by
setting $\L=0$ and taking the limit \UV.

Given the evolution equation in $\L$, the vertices of
$\G[A,\psi,\bar\psi;\L]$ are determined by supplementing
appropriate boundary conditions.
As we shall discuss in this paper this is the place in which the gauge
symmetry properties must be implemented. The boundary conditions for
the various vertices in  $\G[A,\psi,\bar\psi;\L]$ depend
on their dimension in mass.
Vertices with non negative mass dimension are called ``relevant'',
while the others are called ``irrelevant''.
Notice that they are irrelevant from the point of view of dimensional
counting but they actually describe the interaction.
As we shall see the relevant vertices reduce to seven couplings.
One assumes the following boundary conditions:
\par\noindent
1) at $\L=\L_0$ all ``irrelevant'' vertices vanish.
This corresponds to assume that, at the UV scale, the
vertices with negative mass dimension become proportional
to inverse powers of $\L_0$;
\par\noindent
2) at the physical value $\L=0$ the ``relevant'' part of
the effective action $\G_{rel}[A,\psi,\bar\psi;\L=0]$
is given by the classical action, \eg in the Feynman gauge
\beq\nome{action}
\eqalign{
& \G_{rel}[A_\mu,\psi,\bar\psi;\L=0]=
\int_p \; \left \{
\half A_\mu(-p)\, p^2\delta_{\mu\nu}\, A_\nu(p)
\; + i\bar\psi(-p)(\ds p +m)\psi(p)
\right\}
\cr &
+\; ie \, \int_p \int_q\; \bar\psi(-p) \ds A(q) \psi(p-q)
\,,
\;\;\;\;\;\;\;\;\;
\int_p\equiv \int\frac{d^4p}{(2\pi)^4}\;.
}
\eeq
In this way one fixes the seven relevant couplings at $\L=0$
to be the physical mass and charge, and to satisfy Ward identities.

The main result of this paper is the proof, in perturbation theory,
that the functional $\G[A,\psi,\bar\psi;\L]$ obtained from the
renormalization group equation in $\L$ and the above boundary conditions,
satisfies Ward identities in the physical limit \UV and $\L=0$.

The loop expansion for the vertices of $\G[A,\psi,\bar\psi;\L]$
are deduced by solving iteratively the evolution equation in $\L$.
They are given by the usual Feynman diagrams in which all propagators
have Euclidean momentum $p$ in the range $\L^2<p^2<\L_0^2$.
The subtractions needed in order to take the limit \UV are automatically
generated in the process of fixing the physical conditions in
\re{action} for the ``relevant'' vertices.
The fact that Ward identities are satisfied in the physical limit \UV
and $\L=0$ is deeply connected to the property of perturbative
renormalizability, namely to the fact that the divergences in the non
subtracted contributions of Feynman diagrams affect only the relevant
vertices. Thus UV divergences are tamed by imposing the boundary
conditions which generate the necessary subtractions.
The perturbative proof of Ward identities is then obtained by showing that
at $\L=0$ the non subtracted vertices  violate these identities for
\UV only for ``relevant'' contributions which are automatically
cancelled by imposing the physical conditions.
For a previous analysis see for instance \cite{KK,Hu}.

The paper is organized as follows.
First in Sect.~2 we deduce the evolution equation for the
functional $\G[A,\psi,\bar\psi;\L]$ and discuss in details
the boundary conditions.
In Sect.~3 we describe how the loop expansion is obtained
from the iterative solution of the evolution equation for
$\G[A,\psi,\bar\psi;\L]$. We explicitly compute to one
loop order all the vertices together with the axial anomaly
diagram.
To this order we show that the limit \UV can be taken,
due to the proper subtractions, and that Ward identities are
satisfied in the limits \UV and \IR.
The proof to all perturbative order of these identities is
presented in Sect.~4. In Sect.~5 we formulate the proof of
perturbative renormalizability.
The last section contains some remarks and final comments.

\section{Renormalization group flow and effective action}

In order to compute the vertices of the effective action one
needs a regularization procedure of the ultraviolet divergences.
We regularize these divergences by assuming that in the path integral
one integrates only the fields with frequencies smaller than a given
UV cutoff $\L_0$. This procedure is equivalent to assume that the free
photon and electron propagators vanish for $p^2 > \L_0^2$.
The physical theory is obtained by considering the limit \UV.
In order to study the Wilson renormalization group flow \cite{W,P},
one introduces in the free propagators also an infrared (IR) cutoff $\L$.
The cutoff propagator for the electron and photon
(in the Feynman gauge) are
$$
S_{\L\L_0}(p)= \frac{-i K_{\L\L_0}(p) } {\ds p +m} \,,
\;\;\;\;\;
\delta_{\mu\nu} D_{\L\L_0}(p)=
\frac{ \delta_{\mu\nu} K_{\L\L_0}(p)}{p^2} \,,
$$
where $K_{\L\L_0}(p)=1$ in the region
$\L^2 \, \ltap \,p^2\, \ltap \, \L_0^2$ and rapidly vanishing outside.
The corresponding free action is
\beq\nome{0ac}
S_0^{\L,\L_0}= \int_p  \;
\left\{
\half A_\mu(-p)\, p^2\delta_{\mu\nu} \, A_\nu(p)
\;+\;
i\bar\psi(-p)(\ds p +m)\psi(p)
\right\} \, K^{-1}_{\L\L_0}(p)\,.
\eeq
The introduction of a cutoff in the propagators breaks the gauge
invariance properties of the action.
Therefore at the UV scale the interaction Lagrangian must contain all
relevant couplings
with non negative dimensions
\beq\nome{intac}
\eqalign{
S_{int}=
&
\int_p \;
\half A_\mu(-p)\left[(\s_{m_A}^B+p^2\s_\alpha^B )\delta_{\mu\nu}
+p^2\s_A^Bt_{\mu\nu}(p)\right] A_\nu(p)
\;+\; \frac{\s_4^B}{8}\int d^4x (A^2(x))^2
\cr&
+ \int_p \;
i\bar\psi(-p) \left[ \s_{m_\psi}^B +\s_\psi^B (\ds p +m) \right] \psi(p)
\; + \; \int_p \int_{q} \; i\s_e^B \bar\psi(-p)\,\ds A (q)\,\psi(p+q)
\, ,
}
\eeq
where
$$
t_{\mu\nu}(p)\equiv \delta_{\mu\nu}\;-\;\frac {p_\mu p_\nu}{p^2} \,.
$$
The photon and electron mass counterterms, $\s_{m_A}^B$ and $\s_{m_\psi}^B$,
have positive dimensions, while the other parameters are dimensionless.
$\s_A^B$ and $\s_\psi^B$ are related to the wave function renormalizations,
$\s_\alpha^B$ to the gauge fixing parameter renormalization,
$\s_e^B$ is related to the bare charge and $\s_4^B$ is the four photon
interaction coupling.
The complete Lagrangian $S^{\L,\L_0}= S_0^{\L,\L_0}+ S_{int}$
violates the Ward identities, but the couplings $\s_i^B$
should be related in such a way that the (physical) effective
action satisfies them.

In order to simplify the formulae we introduce the compact notation
\beq\nome{source}
\eqalign{&
\Phi_a=(A_\mu,\,\psi_\alpha,\,\bar\psi_\beta)\, ,
\;\;\;\;\;
J_a=(j_\mu,\, i \bar \chi_\alpha,\, -i \chi_\beta)\,,
\cr &
(J,\Phi) \equiv \int_pj_\mu(-p)A_\mu(p)+i \bar \chi(-p)\psi(p)
+i \bar \psi(-p)\chi(p) \,.
}
\eeq
The free cutoff propagators are described by the matrix $D_{ab}$
defined by
$$
S^{\L,\L_0}_0 =
\int_p\half \Phi_a(-p) D^{-1}_{ab}(p;\L)\Phi_b(p) \,,
$$
with $S^{\L,\L_0}_0$ given by \re{0ac}.
The generating functional computed from the cutoff action
$S^{\L,\L_0}$ will depend on $\L_0$ and $\L$. We have
$$
Z[J;\L]=\exp W[J;\L] =\,\int D \left[ \Phi_a \right]\;
\exp \left\{ -S^{\L,\L_0}+ (J,\Phi) \right\}\,
$$
where in $Z[J;\L]$  and $D_{ab}(p;\L)$ we have
explicitly written only the cutoff
$\L$ since we will consider in any case the limit \UV.
The physical functional $Z[J]$ is obtained by taking the limits \IR
and \UV. In these limits the ``bare'' parameters in \re{intac} have to be
fixed in such a way that the ``relevant'' part of the effective
action is given by \re{action}.

\subsection{Evolution equation}
Taking into account that all the $\L$ dependence of $Z[J;\L]$ is
coming from the cutoff in the propagators one easily derives \cite{P}
an evolution equation in $\L$
$$
\dL Z[J;\L]=-\half (2\pi)^8\; \int_q\;\DL{D^{-1}_{ba}(q;\L)}
\frac {\delta^2 Z[J;\L] }{\delta J_b(q) \delta J_a(-q) } \, .
$$
This equation can be converted into an equation for the corresponding
cutoff effective action $\G[\Phi;\L]$ defined as the Legendre
transformation of $W[J;\L]$
$$
\G[\Phi;\L]=-W[J;\L] + W[0;\L]+ (J,\Phi) \, ,
\;\;\;\;\,\;\;
\Phi_a(p) = (2\pi)^4  \frac {\delta W[J;\L] }{\delta J_a(-p)} \, .
$$
This can be done by isolating in $\G[\Phi;\L]$ the contribution of the
two point function
\beq\nome{inv2}
(2\pi)^8\frac {\delta^2\G[\Phi]} {\delta\Phi_a(q)\delta\Phi_b(q')}
=
(2\pi)^4\delta^4(q+q') \, \Delta^{-1}_{ba}(q;\L)+
\G_{ba}^{int}[q',q;\Phi]\,,
\eeq
where $\Delta_{ab}(q;\L)$ is the full propagator.
In Appendix A we derive the evolution equation for the cutoff
effective action and obtain
\beq\nome{eveq}
\dL \biggr\{ \G[\Phi;\L]
-\half \int_q \Phi_a(-q)\, D_{ab}^{-1}(q;\L)\,\Phi_b(q) \biggr\}
=
- \half \int_q M_{ba}(q;\L)\; \bG_{ab}[-q,q;\Phi;\L]\, ,
\eeq
where
\beq\nome{Mab}
M_{ba}=\Delta_{bc}(q;\L) \DL{ D_{cd}^{-1}(q;\L)} \Delta_{da}(q;\L)
\eeq
and $ \bG_{ab}[q,q';\Phi]\,$ is the auxiliary functional defined by the
integral equation (see Fig.~1)
\beq \nome{bG}
\bG_{ab}[q,q';\Phi] = (-)^{\delta_b}\,\G_{ab}^{int}[q,q';\Phi]\,-
\int_{q''} \bG_{cb}[-q'',q';\Phi]
\Delta_{dc}(q'';\L)
\G_{ad}^{int}[q,q'';\Phi]
  \, ,
\eeq
where $\delta_b$ is one if $b$ is a fermion index and zero otherwise.
In terms of the proper vertices
of $\G[\Phi;\L]$ the evolution equations are $(n\ge 3)$
\beq\nome{ceveq}
\dL \G_{c_1\cdots c_n}(p_1, \cdots p_{n};\L)
=  -\half \int_q M_{ba}(q;\L)\;
\bG_{ab,c_1\cdots c_n}(-q,q;p_1, \cdots p_{n};\L) \,,
\eeq
where $\bG_{ab,c_1\cdots c_n}(q,q';p_1, \cdots p_{n};\L)$
are the vertices of the auxiliary functional $\bG[\Phi;\L]$.
For the two point function we write
$$
\Delta^{-1}_{ab}(q;\L)=D^{-1}_{ab}(q;\L)+\Pi_{ab}(q;\L)
$$
and one has
\beq\nome{peveq}
\dL \Pi_{cc'}(p;\L) =  -\half \int_q M_{ba}(q;\L)\;
\bG_{ab,cc'}(-q,q;-p,p;\L) \,.
\eeq
The vertices of the auxiliary functional are obtained
in terms of the proper vertices by expanding \re{bG} and one finds
\beq\nome{bGc}
\eqalign{
&
\bG_{ab, c_1\cdots c_n} (q,q';p_1,\cdots p_{n};\L)=
 \G_{a b c_1\cdots c_n} (q,q',p_1,\cdots p_{n};\L)
\cr&
-{\sum}'
\G_{a c_{i_1}\cdots c_{i_{k}}c'} (q,p_{i_1},\cdots,p_{i_{k}},Q;\L)
\;\Delta_{c'c''}(Q;\L)
\bG_{c''b,c_{i_{k+1}}\cdots c_{i_{n}}}
(-Q,q';p_{i_{k+1}},\cdots p_{i_{n}};\L)
}
\eeq
where $Q=q+p_{i_1}+\cdots p_{i_{k}}$,
and $\sum'$ is the sum over the combinations of photon and
fermion indices $(i_1 \cdots i_{n})$ taking properly into account the
symmetrization and anti-symmetrization.

\subsection{Boundary conditions}
The evolution equation for $\G[\Phi;\L]$ has to be supplemented by the
appropriated boundary conditions for the relevant couplings at $\L=0$
and for the ``irrelevant'' vertices at $\L=\L_0$.
The relevant part of the effective action can be written in
terms of the seven parameters as follows
\beq\nome{rel}
\eqalign{
&
\G_{rel}[\Phi;\L]=
\half \int_p \Phi_a(-p)\,D^{-1}_{ab}(p;\L)\, \Phi_b(p) \;+\int_p \;
i\bar\psi(-p) \left[ \s_{m_\psi}(\L) +\s_\psi(\L) (\ds p +m) \right] \psi(p)
\cr&
+\int_p \;
\half A_\mu(-p)\left[(\s_{m_A}(\L) +p^2\s_{\alpha}(\L) )\delta_{\mu\nu}
+p^2\s_A(\L)t_{\mu\nu}(p)\right] A_\nu(p)
\cr&
+ \int_p \int_{q} \; i\s_e(\L) \bar\psi(-p)\,\ds A (q)\,\psi(p+q)
\;+\; \frac{\s_4(\L)}{8}\int d^4x (A^2(x))^2
\, .
}
\eeq
As described in the Introduction, the boundary conditions are:

\noindent
1) at $\L=0$
\beq\nome{bc1}
\s_e(0) =e \,,\;\;\;\;\;
\s_i(0) =0 \,,
\;\;\;\;\;\mbox{for} \; i=m_A\,,m_\psi\,,\alpha\,,A\,,\psi\,,4;
\eeq
2) at $\L=\L_0$
\beq\nome{bc2}
\G_{irrel}[\Phi;\L=\L_0]=0\,,
\eeq
with $\G=\G_{rel}+\G_{irrel}$.
Ward identities are satisfied then at $\L=0$  for the relevant part
of the effective action.
All couplings with negative dimension entering in $\G_{irrel}[\Phi;\L]$
can be neglected for large $\L$.

We analyse in detail these boundary conditions.

\noindent
1) The photon propagator has the form
\beq\nome{propp}
\Delta^{-1}_{\mu\nu}(p;\L)= \delta_{\mu\nu}\, p^2 \,K_{\L\L_0}^{-1}(p)
+\Pi_{\mu\nu}(p;\L)\,,
\;\;\;\;\;
\Pi_{\mu\nu}(p;\L)=\delta_{\mu\nu}\Pi_L(p;\L)
+t_{\mu\nu}(p)\,\Pi_T(p;\L)
\eeq
with the longitudinal and transverse components given by
$$
\Pi_L(p;\L)=\frac{p_\mu p_\nu}{p^2}\Pi_{\mu\nu}(p;\L)\,,
\;\;\;\;\;
\Pi_T(p;\L)=\third
\left(\delta_{\mu\nu}-4\frac{p_\mu p_\nu}{p^2}\right) \Pi_{\mu\nu}(p;\L)\,.
$$
Using the three relevant couplings in \re{rel} we have
\beq
\eqalign{
\Pi_L(p;\L)=\s_{m_A} & (\L)+p^2 \s_\alpha(\L) + \Sigma_L(p;\L)\,,
\;\;\;\;\;\;\;\;
\Pi_T(p;\L)=p^2 \s_A(\L) + \Sigma_T(p;\L)\,, \cr
& \Sigma_{T,L}(0;\L)=0\,,\;\;\;\;\;\;\;\;
\frac{\partial \Sigma_{T,L}(p;\L)}{\partial p^2}=0
\;\;\;\; \mbox{at} \;\;\;\; p^2=\mu^2\,,
}
\eeq
where $\mu$ is a subtraction point.
{}From this definition we can factorize in the vertices $\Sigma_{T,L}$
a dimensional function of $p$, thus they are ``irrelevant''.
The boundary conditions are then
\beq\nome{bc3}
\s_{m_A}(0)=\s_\alpha(0)=\s_A(0)=0\;,
\;\;\;\;\;\;\;\; \Sigma_{T,L}(p;\L_0)=0\,.
\eeq
2) The fermion propagator has the form
\beq\nome{prope}
\eqalign{
S^{-1}(p;\L) & =
i(\ds p +m)K^{-1}_{\L\L_0}(p) + \Pi(p;\L)\,,
\;\;\;\;\;\;
\Pi(p;\L)=i\s_{m_\psi}(\L)+i(\ds{p}+m) \s_\psi(\L)+\Sigma(p;\L)\,,
\cr
& \Sigma(p;\L)=0 \;\;\;\;\;\mbox{at}\;\;\;\; \ds{p}=-m \,,
\;\;\;\;\;\;\;\;\;\;
\frac{\partial}{\partial p_\mu}
\Sigma(p;\L)=0 \;\;\;\;\;\mbox{at}\;\;\;\; p=0 \,,
}
\eeq
where $\s_{m_\psi},\s_\psi$ are two relevant couplings and the
vertex $\Sigma(p;\L)$ is irrelevant.
The boundary conditions are then
\beq\nome{bc4}
\s_{m_\psi}(0)=\s_\psi(0)=0\,,\;\;\;\;\;\;\;\;\Sigma(p;\L_0)=0\,.
\eeq
3) For the electron-photon vertex we have one relevant coupling
$$
\G_\mu(p,p';\L)=i \g_\mu \, \s_e(\L) +\S_\mu(p,p';\L)\,,
\;\;\;\;\;\;\;\;\;\;
\S_\mu(0,0;\L)=0
$$
and the boundary conditions are
\beq\nome{bc5}
\s_e(0)=e\,,\;\;\;\;\;\;\;\;\S_\mu(p,p';\L_0)=0\,.
\eeq
4) For the four photon interaction vertex one has
\beq
\eqalign{
\G_{\mu_1 \cdots\mu_4}(p_1 \cdots p_4;\L)=
\s_4(\L) ( &
\delta_{\mu_1 \mu_2}\delta_{\mu_3 \mu_4}+
\delta_{\mu_1 \mu_3}\delta_{\mu_2 \mu_4}+
\delta_{\mu_1 \mu_4}\delta_{\mu_2 \mu_3}
)
+\S_{\mu_1 \cdots \mu_4}(p_1 \cdots p_4;\L)\,,
\cr
& \S_{\mu_1 \cdots \mu_4}(0,0,0,0;\L)=0
}
\eeq
and the boundary conditions are
\beq\nome{bc6}
\s_4(0)=0\,,\;\;\;\;\;\;\;\;\S_{\mu_1 \cdots \mu_4}(p_1 \cdots p_4;\L_0)=0\,.
\eeq
5) For all other vertex functions the dimensions are negative
and one has
\beq\nome{bc7}
\G_{c_1 \cdots c_n}(p_1 \cdots p_n;\L_0)=0\,.
\eeq
Finally we observe that the matrix $M_{ab}(q;\l)$ in \re{Mab}
for the fermion and photon case
can be written in the following form
\beq
\nome{Mab'}
\eqalign{
M_{\alpha\beta}(q;\l)=&
-\left[1+S_{\L\L_0}(q)\,\Pi(q;\l)\right]^{-2}
\frac{-i}{\ds{q}+m}
\l\partial_\l K_{\l\L_0}(q)\,,
\cr
M_{\mu\nu}(q;\l)=&
-\biggr\{
\left[1+ D_{\L\L_0}(q)\,\Pi_L(q;\l)\right]^{-2}
\frac{q_\mu q_\nu}{q^2} \cr &
+\,\left[ 1+D_{\L\L_0}(q)\,(\Pi_L(q;\l)+\Pi_T(q;\l))\right]^{-2}
\, t_{\mu\nu}(q)
\biggr\}
\frac{1}{q^2}\l\partial_\l K_{\l\L_0}(q)\, ,
}
\eeq
where we used the propagators decompositions in \re{propp} and
\re{prope}. Notice that $\l\partial_\l K_{\l\L_0}(q)$ is different
from zero only for $q^2 \simeq \l^2$, and is then independent of $\L_0$.

\section{Loop expansion}
The loop expansion is obtained by solving iteratively \re{ceveq}
and \re{peveq}.
Here we compute the first loop. The starting point of the iteration  is
the zero loop order in which,
from \re{bc1}, the only non vanishing coupling is
$$
\s_e^{(0)}(\L)= e\,.
$$
The zero loop auxiliary vertex corresponding to the electron
emitting $n$ photons is given by Fig.~2a
\beq\nome{bg0}
\bG^{(0)}_{\alpha\beta, \mu_1 \cdots \mu_n}(q,q';p_1 \cdots p_n;\L)=
i(-e)^n \left\{
I^{(\L\L_0)}_{\mu_1\cdots \mu_n}(q;p_1,\cdots p_n)
\right\}_{\alpha\beta} +\mbox{permutations}\,,
\eeq
where $q'$ is given by momentum conservation,
\beq\nome{IK}
\eqalign{
&
I^{(\L\L_0)}_{\mu_1\cdots \mu_n}(q;p_1,\cdots p_n)\equiv
I_{\mu_1\cdots \mu_n}(q;p_1,\cdots p_n)
K_{\L\L_0}(q+p_1,\cdots,q+p_1+\cdots p_{n-1})\,,
\cr &
I_{\mu_1\cdots \mu_n}(q;p_1,\cdots p_n)=
\g_{\mu_1}
\frac {1} {\ds q + \ds p_1 +m }\g_{\mu_2}\cdots
\frac {1} {\ds q + \ds p_1+\cdots  \ds p_{n-1} +m }
\g_{\mu_n}
}
\eeq
and we have introduced the general cutoff function
$$
K_{\L\L_0}(q_1,q_2,\cdots,q_n) \equiv
K_{\L\L_0}(q_1)K_{\L\L_0}(q_2)\cdots
K_{\L\L_0}(q_n)
\,.
$$
The other zero loop auxiliary vertices are written in a similar way.

The one loop vertices are obtained by using in the right
hand side of \re{eveq} the auxiliary vertices and propagators at zero loop.
Once the boundary conditions are imposed, the one loop vertices are given
in terms of subtracted integrands. Then all momentum integrations
are UV convergent and we can take the limit \UV.
We show also that at $\L=0$ and \UV the Ward identities
are satisfied to this order.

\vskip .3 true cm
\noindent
1. {\it Electron propagator}
\vskip .2 true cm

The one loop contribution for the electron propagator is obtained from
the evolution equation \re{peveq} by using in the r.h.s. the zero loop
contribution
\beq\nome{S1}
\dL \Pi^{(1)}_{\alpha\beta}(p;\L)=
-\half \int_q \, M^{(0)}_{ba}(q;\L)\,
\bG^{(0)}_{ab, \alpha \beta}(-q,q;-p,p;\L)\, ,
\eeq
where $M^{(0)}_{ba}=-\DL D_{ba}$ and the auxiliary
vertices
$\bG^{(0)}_{\alpha' \beta',\alpha \beta }$ and
$\bG^{(0)}_{\mu   \nu,   \alpha \beta   }$
are given in Fig.~2b. By noticing that there is a factor $2$ for both cases
coming from the permutations of the vertices in $\bG$ (see Eq.\re{bGc}),
the solution of Eq.~\re{S1} is
$$
\Pi'(p;\L)=ie^2\int_q \frac 1 {q^2}
\g_\rho \frac{1}{\ds p+\ds q +m} \g_\rho
K_{\L\L_0}(q,q+p)\,,
$$
apart from a term constant in $\L$. This is fixed by imposing the
boundary conditions \re{bc4} and we obtain the one loop contribution
$$
\Pi^{(1)}(p;\L) =
\Pi'(p;\L)-\Pi'(\bp;0)-
(p_\mu-\frac m 4 \g_\mu)\partial_{p'_\mu}\Pi'(p';0)|_{p'=0}
\,,\;\;\;\;\;\;\ds{\bp}=-m\,.
$$
Because of these subtractions, we can take the limit \UV
\beq\nome{pi1}
\eqalign{
\Pi^{(1)}(p;\L)=
&
ie^2\int_q
\frac 1 {q^2}
\left\{
\g_\rho \frac{1}{\ds p+\ds q +m} \g_\rho
-
\g_\rho \frac{1}{\ds \bp+\ds q +m} \g_\rho
+(p_\mu-\frac m 4 \g_\mu)\, \chi_\mu(0,0,q)
\right\}
\cr&
+ ie^2\int_q
\frac 1 {q^2}
\g_\rho \frac{1}{\ds p+\ds q +m} \g_\rho
\;(K_{\L\infty}(q,q+p)-1) \,,
}
\eeq
where
$$
\chi_\mu(p,p',q)\equiv
\g_\rho
\frac 1 {\ds q +\ds p + m} \g_\mu \frac 1 {\ds q +\ds p' +m}
\g_\rho\,.
$$
The last term, which is $\L$-dependent, is convergent due to
the finite range of integration coming from the IR cutoff.
As we shall see this term is violating the Ward identities but
is vanishing at the physical value $\L=0$ since
$K_{\L\infty}(p)\to 1$ for \IR.

By using a sharp momentum cutoff in \re{pi1}, we find that
for $\L\to \infty$ the relevant couplings (\ie the
parameters in the bare Lagrangian) are given by
\beq\nome{bare1}
\s_{m_\psi}^{(1)}(\L)=\frac{3e^2}{16\pi^2}m\log\frac{\L^2}{m^2}
+\cal{O}(1)
\,,\;\;\;\;\;\;\;\;
\s_{\psi}^{(1)}(\L)=\frac{e^2}{16\pi^2}\log\frac{\L^2}{m^2}
+\cal{O}(1)\,.
\eeq
In the following we will compute the other bare parameters by
using the same cutoff function.

\vskip .3 true cm
\noindent
2. {\it Electron vertex}
\vskip .2 true cm

For this vertex one proceeds as before. The evolution equation
\re{ceveq} and the conditions \re{bc5} give
$$
\G^{(1)}_\mu(p,p';\L)=\G'_\mu(p,p';\L)-\G'_\mu(0,0;0)\,,
$$
$$
\G'_\mu(p,p';\L)=-ie^3\int_q\frac 1 {q^2} \chi_\mu(p,p',q)
K_{\L\L_0}(q,q+p,q+p')\,.
$$
By taking the limit \UV we find
\beq
\eqalign{
\G^{(1)}_\mu(p,p';\L)=
&-ie^3\int_q\frac 1 {q^2}
\left\{ \chi_\mu(p,p',q) - \chi_\mu(0,0,q) \right\}
\cr &
-ie^3\int_q\frac 1 {q^2}
\chi_\mu(p,p',q)\left\{ K_{\L\infty}(q,q+p,q+p')-1 \right\}\,.
}
\eeq
As before the last term is violating Ward identities but is vanishing
for $\L\to 0$.

For large values of $\L$ we find
\beq\nome{bare2}
\s_{e}^{(1)}(\L)=\frac{e^3}{16\pi^2}\log\frac{\L^2}{m^2}
+\cal{O}(1)\,.
\eeq
Notice that at the leading order in $\L$ we have $\s_e^{(1)}=\s_\psi^{(1)}$,
however this relation is violated by the finite parts.
As anticipated the action at the UV scale is not gauge invariant.

\vskip .3 true cm
\noindent
3. {\it Ward identity for electron propagator and vertex}
\vskip .2 true cm

The fact that at $\L=0$ the electron propagator and electron-photon vertex
satisfy the Ward identity can be shown by using the identity
$$
(p'-p)_\mu\, \chi_\mu(p,p',q)=
\g_\rho\left(
\frac 1 {\ds q +\ds p +m} -\frac 1 {\ds q +\ds {p'} +m}
\right)\g_\rho \,.
$$
{}From this we obtain
\beq\nome{WIvv}
\eqalign{
&
(p'-p)_\mu \G^{(1)}_\mu(p,p';\L)
-e\left( \Pi^{(1)}(p';\L)-\Pi^{(1)}(p;\L) \right)
\cr&
=
-ie^3\int_q \frac 1 {q^2}
\g_\rho
\frac {K_{\L\infty}(q,p+q)} {\ds q + \ds p +m}
\g_\rho
\left[ K_{\L\infty}(p'+q)-1 \right]
-\;\{ p \to p' \} \,.
}
\eeq
For $\L=0$ the r.h.s. vanishes thus the Ward identity is satisfied.
Notice that for $\L\neq 0$ and finite the violation is not a
polynomial in $p$ and $p'$ and therefore receives contribution both
from the relevant and the irrelevant couplings.

\vskip .3 true cm
\noindent
4. {\it Photon propagator and Ward identity}
\vskip .2 true cm

For the longitudinal and the transverse part of the photon propagator at one
loop  the evolution equation
\re{ceveq} and the boundary conditions \re{bc3} give
\beq\nome{P1'}
\eqalign{&
\Pi^{(1)}_{L,T}(p;\L)=\Pi'_{L,T}(p;\L)-\Pi'_{L,T}(0;0)
-p^2\,\frac{\partial}{\partial\bp^2}\Pi'_{L,T}(\bp;0)\,,
\;\;\;\;\;\;\;\; \bp^2=\mu^2\,,
\cr &
\Pi'_{L,T}(p;\L)=e^2 \int_q I_{L,T}(q;p)K_{\L\L_0}(q,q+p) \,,
}
\eeq
with
\beq
\eqalign{
&I_L(q;p)=
Tr\left(\frac 1 {\ds q+m} \,\frac{p_\mu p_\nu}{p^2} I_{\mu\nu}(q;p,-p)
\right)
\,,
\cr
&I_T(q;p)= \frac 1 {3}\left(\delta_{\mu\nu}-4
\frac{p_\mu p_\nu}{p^2}\right)
Tr\left(\frac 1 {\ds q+m}I_{\mu\nu}(q;p,-p)\right)
\,,
}
\eeq
where $I_{\mu\nu}$ is defined in \re{IK} and we have $\Pi'_T(0;0)=0$.
By taking the limit \UV we find
$$
\Pi_{L,T}^{(1)}(p;\L)=
e^2\int_q I^s_{L,T}(q;p)
+e^2\int_q I_{L,T}(q;p)
\left( K_{\L\infty}(q,q+p) -1\right)\,,
$$
where
$I^s_{L,T}$ are subtracted integrands
$$
I^{s}_{L,T}(q;p)= I_{L,T}(q;p)- I_{L,T}(q;0)-
p^2\,\frac{\partial}{\partial\bp^2}I_{L,T}(q;\bp)\,.
$$
For large $\L$ the relevant couplings are given by
\beq\nome{bare3}
\s_{m_A}^{(1)}(\L)=\frac{e^2}{8\pi^2}\L^2+\cal{O}(1)\,,\;\;\;\;
\s_{\alpha}^{(1)}(\L)=
\cal{O}(\frac{1}{\L^2})\,,\;\;\;\;
\s_{A}^{(1)}(\L)=-\frac{e^2}{12\pi^2}\log\frac{\L^2}{m^2}
+\cal{O}(1)\,.
\eeq

We now show that to one loop order for $\L=0$ the physical propagator
is purely transverse ($\Pi^{(1)}_L(p;0)=0$).
To show this, in Eq.~\re{P1'} one writes
$$
I_L(q;p)=\frac 1 {p^2} Tr
\left(\ds p  \frac 1 {\ds q +m}- \ds p \frac 1 {\ds q +\ds p +m} \right)
\,,
$$
changes integration variable in the second term and obtains
$$
\Pi'_L(p;0)= e^2 \frac 1 {p^2} \int_q Tr
\left(\ds p \frac 1 {\ds q +m}\right) \,
\left(K_{0\L_0}(q,q+p)- K_{0\L_0}(q,q-p) \right) \,.
$$
Due to the difference of the two cutoff functions we have that $q^2$
is forced in the region $q^2 \sim \L_0$.
By taking for instance an exponential UV cutoff one has
\beq\nome{Kdiff}
K_{0\L_0}(q+p)- K_{0\L_0}(q-p)=-4\frac{p\cdot q}{\L_0^2}
\left\{1-\frac{p^2}{\L_0^2}+\frac 2 {3}
\frac{(p\cdot q)^2}{\L_0^4} +\cdots\right\}
\, e^{-{q^2}/{\L_0^2}}
\eeq
and obtains
$$
\Pi'_L(p;0)=  -\frac{e^2}{16\pi^2}(\L_0^2-2m^2)\,
+\,\frac{e^2}{24\pi^2}p^2 + {\cal O} (\frac{m^2, p^2}{\L^2_0})\,.
$$
In this calculation the effect of the non invariant regularization
is clear.
The divergent integral with the cutoff functions gives a surface
term which destroys the transversality of the propagator.
However the longitudinal contributions are of relevant type, thus they
are cancelled by imposing the boundary conditions. One finds
$$
\Pi^{(1)}_L(p;0)= \Pi'_L(p;0)- \Pi'_L(0;0)
-p^2 \partial_{p'^2} \Pi'_L(p';0)|_{p'^2=\mu^2}=
{\cal O} (\frac{m^2, p^2}{\L^2_0}) \to 0\,.
$$
For $\L\neq 0$ the longitudinal part of the photon propagator
is different from zero both in the relevant ($\s_{m_A}(\L)$ and
$\s_\alpha(\L)$) and irrelevant ($\S_L(p;\L)$) parts.

\vskip .3 true cm
\noindent
5. {\it Multi-photon vertices}
\vskip .2 true cm

The evolution equation and the boundary conditions \re{bc6} and \re{bc7}
give
$$
\G^{(1)}_{\mu_1\cdots \mu_n}(p_1,\cdots p_n;\L)=
\G'_{\mu_1\cdots \mu_n}(p_1,\cdots p_n;\L)
-\G'_{\mu_1\cdots \mu_n}(0,\cdots 0;0)\,\delta_{n,4}\,,
$$
where the subtraction is required only for the four photon vertex. The non
subtracted vertices are
\beq\nome{F1}
\G'_{\mu_1\cdots \mu_n}(p_1,\cdots p_n;\L)=
(-e)^n \int_q \, Tr\,\left\{
\frac{K_{\L\L_0}(q)}{\ds q +m}
I^{(\L\L_0)}_{\mu_1\cdots \mu_n}(q;p_1,\cdots p_n)\right\}
+\mbox{permutations}
\eeq
where $I^{(\L\L_0)}_{\mu_1\cdots \mu_n}$ is defined in \re{IK}.
For $n> 4$ this integral is convergent and we can remove the
cutoff $\L_0$.
In the case $n=4$ one needs to consider the subtraction
\beq
\eqalign{
&
\G^{(1)}_{\mu_1\cdots \mu_4}(p_1,\cdots p_4;\L)=
\G'_{\mu_1\cdots \mu_4}(p_1,\cdots p_4;\L)
-\G'_{\mu_1\cdots \mu_4}(0,\cdots 0;0)
\cr
&
=e^4 \int_q \, Tr\,\left\{ \frac 1 {\ds q +m} \left(
I_{\mu_1\cdots \mu_4}(q;p_1,\cdots p_4)-
I_{\mu_1\cdots \mu_4}(q;0,\cdots 0)\right)
\right\}
\cr &
+e^4 \int_q \, Tr\,\left\{ \frac 1 {\ds q +m}
I_{\mu_1\cdots \mu_4}(q;p_1,\cdots p_4)\right\}
\left[K_{\L\infty}(q,q+p_1\cdots) -1\right]
+\mbox{permutations}\,.
}
\eeq
The various terms in the non-subtracted vertex $\G'_{\mu_1\cdots \mu_4}$
are logarithmically divergent when removing the cutoff.
However, by summing over the permutations, these divergent terms
cancel, giving a finite result.
In dimensional regularization, at vanishing momenta the vertex is zero.
In the present regularization one finds instead a finite result for
\UV. At $\L=0$ we have
\beq\nome{G4}
\G'_{\mu_1\cdots \mu_4}(0,0,0,0;0)
=
\frac {e^4}{12\pi^2}
\left(
  \delta_{\mu_1\mu_2} \delta_{\mu_3 \mu_4}
+ \delta_{\mu_1\mu_3} \delta_{\mu_2 \mu_4}
+ \delta_{\mu_1\mu_4} \delta_{\mu_2 \mu_3}
\right)\,.
\eeq
The non vanishing of this finite term implies that the subtraction
required by the constraint \re{bc6} is necessary even if its contribution
is finite.
With this subtraction we have $ \G_{\mu_1\cdots \mu_4}(0,\cdots 0;0)=0$
which is essential in order to satisfy the Ward identity.
For large $\L$ the relevant coupling is
\beq\nome{bare4}
\s_4^{(1)}(\L)=-\frac {e^4}{12\pi^2}+{\cal O}(\frac{1}{\L^2}) \,.
\eeq

\vskip .3 true cm
\noindent
6. {\it Ward identities for the multi-photon vertices}
\vskip .2 true cm

The basic identity used to prove the Ward identities is
\beq\nome{basic}
p_{1\mu_1} \frac 1 {\ds q +m} I_{\mu_1\cdots \mu_n}(q;p_1,\cdots p_n)=
\left( \frac 1 {\ds q +m } - \frac 1 {\ds q + \ds p_1 +m } \right)
I_{\mu_2\cdots \mu_n}(q+p_1;p_2,\cdots p_n) \,.
\eeq
By using \re{basic} in \re{F1} and changing integration variable we can write
\beq\nome{div}
\eqalign{
&
p_{1\mu_1}
\G'_{\mu_1\cdots \mu_n}(p_1,\cdots p_n;\L)
=(-e)^n \int_q \, \left[ K_{\L\L_0}(q+p_1)-K_{\L\L_0}(q+p_2)\right]
\cr&
\times Tr\,\left\{ \frac{K_{\L\L_0}(q)}{\ds q +m}
I^{(\L\L_0)}_{\mu_2\mu_{3}\cdots \mu_{n}}(q+p_1;p_2,p_{3}\cdots p_{n})
\right\}
\;\;\;\;\; + \cdots \,,
}
\eeq
where the dots correspond to the sum over the permutations of
$(2,\dots n)$.
Consider first the case $n>4$ which needs not subtractions.
By expanding for $\L=0$ the differences of cutoff functions,
similarly to \re{Kdiff}, we obtain the following behaviour for
large $\L_0$
\beq\nome{WI2}
p_{1\mu_1}
\G^{(1)}_{\mu_1\cdots \mu_n}(p_1,\cdots p_n;0)
=c\,( p_{1\mu_2}\,\delta_{\mu_3\mu_4}\cdots \delta_{\mu_{n-1}\mu_n}
+\mbox{symm.})
\left(\frac 1{\L_0}\right)^{n-4}
+ {\cal O}\left( p_1\frac{P^2}{\L_0^{n-2}} \right)
\to 0\,,
\eeq
where $P$ is a combination of external momenta and $c$ a numerical constant.
Notice that \re{WI2} implies
$$
\G^{(1)}_{\mu_1\cdots \mu_n}(0,\cdots 0;0)=0\,.
$$
In the case $n=4$
one obtains a finite result
$$
p_{1\mu_1} \G'_{\mu_1\cdots \mu_4}(p_1,\cdots p_4;0)
= \frac{e^4}{12\pi^2} \left(
p_{1\mu_2}\delta_{\mu_3\mu_4}+
p_{1\mu_3}\delta_{\mu_2\mu_4}+
p_{1\mu_4}\delta_{\mu_3\mu_2}
\right)
+{\cal O}(p_1\frac{P^2}{\L_0^2})\,,
$$
where the constant term comes from the first term in the $1/\L_0$ expansion
of the differences of cutoffs.
Combining this with the result \re{G4} one finds for $\L=0$ and \UV
that the Ward identity is satisfied to this order
\beq\nome{WI2'}
p_{1\mu_1} \G^{(1)}_{\mu_1\cdots \mu_4}(p_1,\cdots p_4;0)
=\,{\cal O}(p_1\frac{P^2}{\L_0^2}) \to 0 \,.
\eeq

\vskip .3 true cm
\noindent
5. {\it Axial anomaly}
\vskip .2 true cm

Finally we consider the axial Ward identity for
the VVA (axial) $T_{\mu\nu\rho}$ and the VVP
(pseudoscalar) $T_{\nu\rho}$ vertices obtained by
insertion of the axial $i\bar\psi\g_\mu\g_5\psi$
and pseudoscalar current $i\bar\psi\g_5\psi$
respectively
\bminiG{W5}
\nome{W5a}
&
p_\nu\,T_{\mu\nu\rho}(p,p')=0\,,
\;\;\;\;\;\;
p'_\rho\,T_{\mu\nu\rho}(p,p')=0\,,
\\
\nome{W5b}
&
(p+p')_\mu\,T_{\mu\nu\rho}(p,p')=-2m T_{\nu\rho}(p,p')
-\frac{1}{2\pi^2}\eps_{\alpha\nu\rho\beta} p_\alpha p'_\beta\,.
\emini
As well known at one loop the axial Ward identity  develops an
anomalous term.

Here we recover this result in our scheme.
The generating functional for these vertices are obtained
by adding the corresponding source terms in \re{source}.
Since these sources do not couple, no Legendre transformation
is needed and the evolution equation for $T_{\mu\nu\rho}$ and
$T_{\nu\rho}$ has the same structure of \re{ceveq}.
The boundary conditions are fixed according to the same arguments
of relevance previously discussed.
The Lorentz covariance and symmetrization with respect to the two photons
give the following decompositions
\beq\nome{dec}
\eqalign{
&
T_{\mu\nu\rho}(p,p';\L)=
\eps_{\mu\nu\rho\s}
(p-p')_\s\,A(p,p';\L)
+
\eps_{\mu\nu\alpha\beta}
p_\alpha p'_\beta (p+p')_\rho \,\S_1(p,p';\L)
\cr&
+
\eps_{\mu\alpha\rho\beta}
p_\alpha p'_\beta (p+p')_\nu \,\S_1(p',p;\L)
+
\eps_{\alpha\nu\rho\beta}
p_\alpha p'_\beta (p+p')_\mu \,\S_2(p,p';\L)
}
\eeq
and
$$
T_{\mu\nu}(p,p';\L)=
\eps_{\mu\nu\alpha\beta}
p_\alpha p'_\beta \,\S_3(p,p';\L) \,.
$$
{}From dimensional counting we have that only the function $A$ has
non-negative dimension and we can write
\beq\nome{A}
A(p,p';\L)=\s_5(\L)+ \S_5(p,p';\L)\,,
\eeq
where $\s_5(\L)$ is a relevant dimensionless constant while
$\S_5$ is irrelevant since $\S_5(0,0;\L)=0$.
The boundary conditions are
$$
\s_5(0)=0\,,\;\;\;\;\;\;\;\;
\S_i(p,p';\L_0)=0\,, \;\;\;\;\;\; i=1,2,3,5
\,.
$$
The last condition is fixed by requiring that the relevant part of effective
action satisfy the vector identity \re{W5a}.
Imposing these boundary conditions we find, at one loop level
\beq\nome{T1l}
T^{(1)}_{\mu\nu\rho}(p,p';\L)=
T'_{\mu\nu\rho}(p,p';\L)
-\eps_{\mu\nu\rho\s}(p-p')_\s\,
A'(0,0;0)\,,
\eeq
where $T'_{\mu\nu\rho}$ is the one loop graph
\beq\nome{T'1l}
\eqalign{
T'_{\mu\nu\rho}(p,p';\L)=
\int_q Tr
\frac{K_{\L\L_0}(q)}{\ds q +m}
\g_\mu & \g_5
\frac{K_{\L\L_0}(q-p-p')}{\ds q -\ds p -\ds p' +m}
\g_\rho
\frac{K_{\L\L_0}(q-p)}{\ds q -\ds p +m}
\g_\nu
\cr &
+\;\;(p,\nu)\to(p',\rho)\,,
}
\eeq
and
$$
\eps_{\mu\nu\rho\s}\,A'(0,0;\L)=
\frac{\partial}{\partial p_\s}
T'_{\mu\nu\rho}(p,p';\L)|_{p=p'=0}\,.
$$
The calculation of the Ward identities follows the same steps as the
previous ones, in particular we must take into account the surface terms
coming from the difference of cutoff functions. We find at $\L=0$
\bminiG{W5'}
\nome{W5'a}
p_\nu
T'_{\mu\nu\rho}(p,p';0) &=& \frac{1}{6\pi^2}
\eps_{\mu\alpha\rho\beta}p_\alpha p'_\beta+\cal{O}(\frac{1}{\L_0^2})\,,
\\
\nome{W5'b}
p'_\rho
T'_{\mu\nu\rho}(p,p';0) &=& \frac{1}{6\pi^2}
\eps_{\mu\nu\alpha\beta}p_\alpha p'_\beta+\cal{O}(\frac{1}{\L_0^2})\,,
\\
\nome{W5'c}
(p+p')_\mu
T'_{\mu\nu\rho}(p,p';0) &=& -\frac{1}{6\pi^2}
\eps_{\alpha\nu\rho\beta}p_\alpha p'_\beta-
2mT^{(1)}_{\nu\rho}(p,p';0)+\cal{O}(\frac{1}{\L_0^2})\,,
\emini
where $T^{(1)}_{\nu\rho}$ is the one loop VVP vertex
given by the r.h.s. of \re{T'1l}
with $\g_\mu\g_5\to \g_5$ and no subtraction is needed.
Using \re{W5'a} and \re{W5'b} and
a Lorentz decomposition of $T'_{\mu\nu\rho}$ analogous to \re{dec},
we have
\beq\nome{A'}
A'(0,0;0)=-\frac{1}{6\pi^2} +{\cal O}(\frac{1}{\L_0^2})\,.
\eeq
In the limit \UV, the Ward identities \re{W5} are obtained from
equation \re{T1l}, after the insertion of \re{W5'} and \re{A'}.

In conclusion we have assumed the vector Ward identities in \re{W5a}
only for the relevant part of the VVA vertex. We have then proved that
the two Ward identities in \re{W5a}-\re{W5b} are satisfied for the
physical VVA and VVP vertex.
In particular the axial anomaly is generated by the surface
contribution associated to a linear divergent graph.

\section{Ward identities}
In this section we prove that the Ward identities are satisfied
in perturbation theory when we take the physical limit.
Therefore in this section we suppress the index $\L$ which is
set at $\L=0$.
We use the standard technique based on Feynman diagrams.
The important point is the analysis of the role of the cutoff
and the UV limit \UV.
As we shall see the non subtracted vertices violate the Ward identities by
surfaces terms coming, as in the one loop case, from the difference of
cutoff functions \re{Kdiff}.
We show that the violating terms are of relevant type thus they
are cancelled by the subtractions coming by imposing the physical boundary
conditions \re{bc1}.
The presence of the difference of cutoff functions \re{Kdiff} in
non subtracted vertices forces the integration momenta to be of the order
of the UV cutoff $\L_0$.
The evaluation of the integral is then obtained by using the Weinberg
theorem \cite{Wth} for the asymptotic behaviour at large momenta
of the vertices in the integrands.
We recall that this behaviour is given by power counting
arguments, apart from logarithmic  corrections, which are inessential
for our perturbative analysis.

We consider first the fermion vertex identity and then the photon
transversality.

\vskip .3 true cm
\noindent
1. {\it Fermion vertex}
\vskip .2 true cm

Consider the general graphs Fig.~3a for the non subtracted fermion
propagator which gives the contribution
\beq\nome{WI11}
\S'^{{\cal G}}(p)=i(-e)^n\int_{q_1}\cdots \int_{q_n}
(2\pi)^4\delta^4(\sum_1^n q_i)\;
G_{\mu_1\cdots \mu_n}(q_1\cdots q_n)\,
I^{(0\L_0)}_{\mu_1\cdots \mu_n}(p;q_1,q_2\cdots q_{n}) \,,
\eeq
where the function $G_{\mu_1\cdots \mu_n}(q_1\cdots q_n)$ corresponds to the
sum of connected and disconnected Feynman graphs in the upper part of
Fig.~3a. This function includes the photon propagators.

The corresponding graphs for the vertex function
are obtained by inserting in Fig.~3a a photon of momentum
$k=p'-p$ with polarization $\mu$ in all possible ways.
We obtain the two contributions of Fig.~4. From the first one in which
the photon is inserted in the fermion line, we find
\beq\nome{WI13}
\eqalign{
&
k_\mu\;\S'^{{\cal G}}_\mu (p,p')=i(-e)^{n+1}\int_{q_1}\cdots \int_{q_n}
(2\pi)^4\delta^4(\sum_1^n q_i)\;
G_{\mu_1\cdots \mu_n}(q_1\cdots q_n)
\cr &
\times
k_\mu\;\sum_{\ell=1}^{n-1}\;
I^{(0\L_0)}_{\mu_1\cdots \mu_{\ell}\mu\cdots \mu_n}
(p;q_1,\cdots q_{\ell},k,\cdots q_{n})\,.
}
\eeq
As we shall see later the contribution of the second type graphs
vanish.

The contributions in \re{WI11} and \re{WI13} are not subtracted
so that we should take $\L_0$ fixed.
The limit \UV can be considered only for the corresponding subtracted
vertices
\beq\nome{WI12}
\eqalign{
&
\S^{\cal G} (p)=\S'^{\cal G} (p)-\S'^{\cal G}
(\bp)-(p_\mu-\frac m 4 \g_\mu)\,
{\partial}_{p'_\mu}\S'^{\cal G} (p')|_{p'=0}\,,
\;\;\;\;\ds{\bp}=-m\,,
\cr&
k_\mu\;\S^{\cal G}_{\mu}(p,p')=k_\mu\;\left(
\S'^{\cal G}_{\mu}(p,p')-\S'^{\cal G}_{\mu}(0,0)\right) \,,
}
\eeq
obtained by imposing the boundary conditions \re{bc4} and \re{bc5}.
Subtraction of subdivergences are considered later.
By using \re{WI11}-\re{WI12} the Ward identity violation
\beq\nome{WI15}
{\cal W}^{\cal G}(p,p';\L_0) =
k_\mu\;\S^{\cal G}_{\mu}(p,p')
-e\left( \S^{\cal G}(p')-\S^{\cal G}(p) \right)
\eeq
becomes
\beq\nome{viol}
{\cal W}^{\cal G}(p,p';\L_0) =
i(-e)^{n+1}\int_{q_1}\cdots\int_{q_n}
(2\pi)^4\delta^4(\sum_1^n q_i)
G_{\mu_1\cdots \mu_n}(q_1\cdots q_n) \,
R^{(\L_0)}_{\mu_1\cdots \mu_n}(p,p',q_1\cdots q_n) \,,
\eeq
where
\beq
\eqalign{
&
R^{(\L_0)}_{\mu_1\cdots \mu_n}(p,p',q_1\cdots q_n)
\cr &
=\sum_{\ell=1}^{n-1}\;k_\mu
\left\{
I^{(0\L_0)}_{\mu_1\cdots \mu_{\ell}\mu\cdots \mu_n}
(p;q_1,\cdots q_{\ell},k,\cdots q_{n})-
I^{(0\L_0)}_{\mu_1\cdots \mu_{\ell}\mu\cdots \mu_n}
(0;q_1,\cdots q_{\ell},0,\cdots q_{n})\right\}
\cr & -
\left(
I^{(0\L_0)}_{\mu_1\cdots \mu_n}(p ;q_1,\cdots q_{n})-
I^{(0\L_0)}_{\mu_1\cdots \mu_n}(p';q_1,\cdots q_{n})
\right)
-k_\mu\;{\partial}_{p''_\mu}\,
I^{(0\L_0)}_{\mu_1\cdots \mu_n}(p'';q_1,\cdots q_{n})|_{p''=0} \;.
}
\eeq
Without the cutoff one has the identity
\beq
\eqalign{
&
\sum_{\ell=1}^{n-1}\;k_\mu\;\left\{
I_{\mu_1\cdots \mu_{\ell}\mu\cdots \mu_n}
(p;q_1,\cdots q_l,k,\cdots q_{n})-
I_{\mu_1\cdots \mu_{\ell}\mu\cdots \mu_n}
(0;q_1,\cdots q_{\ell},0,\cdots q_{n})\right\}
\cr&
=I_{\mu_1\cdots \mu_n}(p ;q_1,\cdots q_{n})
-I_{\mu_1\cdots \mu_n}(p';q_1,\cdots q_{n})
+k_\mu\;{\partial}_{p''_\mu}\,
I_{\mu_1\cdots \mu_n}(p'';q_1, \cdots q_{n})|_{p''=0}\,,
}
\eeq
which gives $R^{(\infty)}=0$.
Actually, since the integration in \re{viol} is divergent,
we should take into account the cutoff functions. We obtain
\beq\nome{RK}
\eqalign{
&
R^{(\L_0)}_{\mu_1\cdots \mu_n}(p,p',q_1\cdots q_n)=
I^{(0\L_0)}_{\mu_1\cdots \mu_n}(p';q_1,\cdots q_{n})
[K_{0\L_0}(p'+q_1)-K_{0\L_0}(p+q_1)]
\cr&
\;\;\;\;\;\;\;\;\;\;\;\;\;\;\;\;\;\;\;\;\;\;\;\;\;\;\;\;\;\;\;\;\;\;\;
+I^{(0\L_0)}_{\mu_1\cdots \mu_n}(p;q_1,\cdots q_{n})
[K_{0\L_0}(p'-q_n)-K_{0\L_0}(p-q_n)]
\cr&
+\sum_{\ell=2}^{n-1}
I^{(0\L_0)}_{\mu_1\cdots \mu_n}(p;q_1,\cdots q_\ell+k,\cdots q_{n})
[K_{0\L_0}(p'+q_1+\cdots +q_{\ell-1})-K_{0\L_0}(p+q_1+\cdots +q_{\ell})]
\cr &
-I_{\mu_1\cdots \mu_n}(0;q_1,\cdots q_{n}) \,
k_\mu\;{\partial}_{p''_\mu}\,K_{0\L_0}(p''+q_1,\cdots p''+q_1+q_{n-1})|_{p''=0}
\,.
}
\eeq
This expression contains differences and derivatives of the cutoff
functions. Thus some of the integration momenta $q_i$ are of order
$Q\sim\L_0$.
We have to distinguish the two cases in which $G_{\mu_1\cdots \mu_n}$
corresponds to connected or disconnected graphs.

When $G_{\mu_1\cdots \mu_n}$ corresponds to connected
graphs, the leading contribution to the violation \re{viol} comes from the
integration region in which all the intermediate photons $(q_1\cdots q_n)$
have large momenta of order $Q$. The resulting
asymptotic behaviour of $R^{(\L_0)}$ is given by
\beq\nome{WI19}
R^{(\L_0)}_{\mu_1\cdots \mu_n}(p,p',q_1\cdots q_n)
\simeq \frac {kQ}{\L^2_0}\, Q^{-n}\,.
\eeq
This comes from the second order of the expansion of \re{RK} in $p$ and $p'$,
namely the first derivatives of $I^{(0\L_0)}$ times the first term of
the expansion of the cutoff differences.
To evaluate the limit for \UV of ${\cal W}^{\cal G}$ we need the asymptotic
behaviour of the connected functions $G_{\mu_1\cdots \mu_n}(q_1\cdots q_n)$
when the momenta become large of order $Q\simeq\L_0$. This is provided by
the Weinberg theorem which gives
$$
G_{\mu_1\cdots \mu_n}(q_1\cdots q_n) \sim Q^{-2n}\,\cdot\,Q^{4-n}\,,
$$
where the first factor comes from the $n$ photon propagators and the second
from the dimensional counting of the vertex with only hard photon lines.
The momentum integration in \re{viol} gives a factor $Q^{4n-4}$, thus the final
asymptotic behaviour for \UV is given by
\beq\nome{viol2}
{\cal W}^{\cal G}(p,p';\L_0)
 \sim \frac{1}{\L_0} \to 0\,.
\eeq
We consider now the case in which the function $G_{\mu_1\cdots \mu_n}$
corresponds to two disconnected graphs, as in Fig.~3b.
We exclude at the moment subgraphs contributing to
fermion self energy. We have again that the leading contribution is
obtained when all $n$ intermediate photon momenta are of order $Q\sim \L_0$.
In fact the Weinberg theorem gives
$G_{\mu_1\cdots \mu_n}\sim Q^4 \cdot Q^{4-3n}$ but the momentum integrations
give now $Q^{-4} \cdot Q^{4n-4}$. Then we obtain the
same result in \re{viol2}. A similar argument holds for the case of any number
of disconnected graphs, excluding fermion self energies.

An independent analysis is needed when $G_{\mu_1\cdots \mu_n}$ contains self
energy corrections to the fermion propagators. This is shown for instance
in Fig.~3c, in which $m$ photon lines belong to the self energy correction.
The leading contribution is obtained when the $m$ photon momenta become large
and we have $G_{\mu_1\cdots \mu_n}\sim Q^{4-3m}$. From the integration we have
$Q^{4-3m}$. However, since $m<n$, the behaviour of the function $R^{(\L_0)}$
is now given by
$$
R^{(\L_0)}_{\mu_1\cdots \mu_n}(p,p',q_1\cdots q_n)
\simeq \frac {kQ}{\L^2_0}\, Q^{-m}\,\cdot\,Q\,.
$$
The reason for the extra power of $Q$, in comparison to \re{WI19}, is that not
all the momenta $q_i$ are large. Thus, when expanding $I^{(0\L_0)}$ in \re{RK},
the derivative with respect to $p$ and $p'$ may act on a fermion propagator
with soft momentum. These behaviours give a non vanishing contribution to
the Ward identity violation ${\cal W}^{\cal G}$. This is expected, since the
self energy correction is a relevant contribution and needs to be subtracted.
Similarly, the insertions of the photon $k$ in the fermion lines among the
$m$ photons of $G'_c$ give vertex corrections,
which are relevant and need subtraction.
All these subtractions restore the behaviour in \re{viol2}, since they force
one to take the derivative with respect to $p$ and $p'$ of the fermion
propagators in the self energy subgraph. In this way one obtains an
additional power of $Q^{-1}$. The result is that also in this case ${\cal
W}^{\cal G}$ is vanishing as in \re{viol2}.

In conclusion, one finds that the differences of the cutoff functions in
$R^{(\L_0)}$ imply that the Ward identity violation is vanishing as an inverse
power of $\L_0$.
Due to this fact, subtractions of the remaining subdivergences,
namely vertex corrections, need not to be analyzed
independently since they diverge only logarithmically in $\L_0$.
This means that we do not need a detailed study of the subdivergences, which
relies on the forest technology \cite{Z}.

The same proof is valid for the general class of Ward identities
relating vertices in which any number of
photons and fermion pairs are emitted from the blob $G$ in Fig.~3a and
Fig.~4.

\vskip .3 true cm
\noindent
2. {\it Photon transversality}
\vskip .2 true cm

We consider the case in which the photon we want to study is emitted by a
fermion loop, as shown in Fig.~5.
We denote by $\sigma$ the spinor and Lorentz indices of the lines in $S$.
The blob corresponds to the sum over all
graphs with their subtractions and includes also disconnected graphs.
If we are dealing with a relevant vertex (that is
$S=\g,3\g,e^+e^-$), we must take into account the overall subtractions,
as required by the boundary conditions.
The case with other external photons
emitted by the same loop will be discussed later.

Associated with the graph in Fig.~5, there are other contributions
coming from the subtractions of subgraphs not contained in the blob.
We analyze first those involving the external photon $p$.
Consider a cut of the diagrams in the blob which disconnects the
$n$ intermediate photons from the external lines $S$.
If the lines which are cut are an electron-positron pair or three photons
(``relevant intermediate lines''), the subgraph identified by the external
photon $p$ and these lines is a relevant vertex, so it must be subtracted.
In order to distinguish these cases, we denote by a box ($B$) the
sum of graphs, with their corresponding subtractions, which do not have
relevant intermediate lines. From this we have that
the graphs in Fig.~5 can be cast in the form of Fig.~6,
\ie the blob can be expanded as a series of boxes.
We denote by $\G'^{b\bx}_{\mu,\s}(p;S)$ the generic term of the expansion
involving $b$ boxes and by $\G^{b\bx}_{\mu,\s}(p;S)$ the vertex obtained
from $\G'^{b\bx}_{\mu,\s}(p;S)$ after having subtracted all the subgraphs
involving $p$.
We will show that $p_\mu\G^{b\bx}_{\mu,\s}(p;S)$ vanishes  as a
negative power of $\L_0$. For this reason, as in the previous case,
the subtractions for other subdivergences (\eg corrections to vertices
and propagators in the fermion loop) need not to be considered
independently, since they are logarithmic in $\L_0$.

In order to  prove the transversality we analyze the asymptotic
behaviour of the intermediate
momentum integrations by using the Weinberg theorem, neglecting all
logarithmic corrections. In particular we consider
the vertices $B_{n\g,me^+e^-}(q_1\cdots;S)$,
corresponding to the general box in Fig.~6 in which enter $n$ photons and
$m$ fermion pairs with hard momentum.
If the state $S$ is ``irrelevant'', that is $S\neq\g,3\g,e^+e^-$, from
the Weinberg theorem one has the asymptotic  behaviour
\beq\nome{box}
B_{n\g,me^+e^-}(q_1\cdots ;S)= {\cal O}(Q^{-n-3m})\;,
\eeq
where all the momenta $q_i$ are large of order $Q$ and the momenta in $S$ are
finite. If $S$ is ``relevant'' one has instead
\beq\nome{box'}
\eqalign{
&B_{n\g,me^+e^-}(q_1\cdots ;S=\g) = {\cal O}(Q^{3-n-3m}) \;,
\cr&
B_{n\g,me^+e^-}(q_1\cdots ;S=3\g) = {\cal O}(Q^{1-n-3m}) \;,
\cr&
B_{n\g,me^+e^-}(q_1\cdots ;S=e^+e^-) = {\cal O}(Q^{1-n-3m})\;.
}
\eeq
{}From these behaviours we immediately deduce the transversality of
$\G^{1\bx}_{\mu,\s}(p;S)$. For the unsubtracted vertices we have
\beq\nome{1box}
\eqalign{
p_\mu \G_{\mu,\s}'^{1\bx}(p;S)=& \int_{q_1} \cdots \int_{q_n}
(2\pi)^4 \delta^4(\sum_1^n q_i +p) \prod_1^n \frac{K_{0\L_0}(q_i)}{q_i^2}
\cr &
\times\, p_\mu \G_{\mu\mu_1\cdots \mu_n}^{(1)}(p,q_1\cdots q_n)
B_{\mu_1\cdots \mu_n}(q_1\cdots q_n;S)\,.
}
\eeq
The fermion loop in Fig.~6 gives rise to the one loop vertex
$\G_{\mu\mu_1\cdots \mu_n}^{(1)}$ with $n+1$
photons; for $n=3$ the subtraction is included.
{}From \re{WI2} and \re{WI2'} we have that the longitudinal part of
$\G_{\mu,\s}'^{1\bx}(p;S)$ comes from the integration region of large $q_i$,
which could compensate the negative power of $\L_0$ in $p\cdot\G^{(1)}$.
In this case we can use the behaviours in \re{box} and \re{box'} and obtain
the following results.

\noindent
a) For the irrelevant vertices the photon is transverse
$$
p_\mu \G_{\mu,\s}^{1\bx}(p;S) = {\cal O}(\frac{1}{\L_0})\to 0\,.
$$
b) For the relevant vertices we find
\beq
\eqalign{
\frac{p_\mu p_\nu}{p^2} \G_{\mu\nu}'^{1\bx}(p)
&= a\L_0^2+a'm^2+bp^2+{\cal O}(\frac{p^4}{\L_0^2}) \,,
\cr
p_\mu \G_{\mu}'^{1\bx}(p,q_1,q_2)
&= c\,\ds p +{\cal O}(p\frac{Q}{\L_0}) \,,
\cr
p_\mu \G_{\mu\mu_1\mu_2\mu_3}'^{1\bx}(p,q_1,q_2,q_3)
&= d\,(p_{\mu_1}\delta_{\mu_2\mu_3}+ \cdots)
+{\cal O}(p\frac{Q^2}{\L_0^2}) \,,
}
\eeq
where $Q$ is a combination of the momenta $p$ and $q_i$.
The constants $a,a', b,c,d$ are relevant couplings which are
cancelled after the overall subtractions leaving longitudinal contributions
which vanish for \UV
\beq\nome{1boxrel}
\eqalign{
\frac{p_\mu p_\nu}{p^2} \G_{\mu\nu}^{1\bx}(p)
&= {\cal O}(\frac{p^4}{\L_0^2}) \to 0 \,,
\cr
p_\mu \G_{\mu}^{1\bx}(q_1,q_2)
&= {\cal O}(p\frac{Q}{\L_0}) \to 0 \,,
\cr
p_\mu \G_{\mu\mu_1\mu_2\mu_3}^{1\bx}(p,q_1,q_2,q_3)
&= {\cal O}(p\frac{Q^2}{\L_0^2}) \to 0 \,.
}
\eeq
We already noticed that the blob of Fig.~5 is not necessarily connected. In
particular we can have some of the $n$ intermediate photons $q_1 \cdots q_n$
disconnected from the external lines $S$. These contributions give for example
corrections to the propagators and vertices in the fermion loop. From the
Weinberg theorem they have the same behaviours as in \re{1boxrel}.

We can now prove by iteration that the behaviour of the diagrams
with $b$ boxes is the same as in the case of one box. From Fig.~6 we
obtain the contribution for $b+1$ boxes in terms of the one with $b$
boxes as follows
\beq\nome{boxes}
\eqalign{
p_\mu \G_{\mu,\s}'^{(b+1)\bx}(p;S) =& \int_{q_1} \int_{q_2}
K_{0\L_0}(q_1)K_{0\L_0}(q_2)(2\pi)^4 \delta^4(q_1+q_2+p)
\cr &
\times
Tr \left\{ p_\mu \G_{\mu}^{b\bx}(p;q_1,q_2)
\frac{-i}{\ds q_1 +m} B_{e^+e^-}(q_1,q_2;S)
\frac{-i}{\ds q_2 +m} \right\}
\cr +&
\int_{q_1} \int_{q_2} \int_{q_3}
(2\pi)^4 \delta^4(q_1+q_2+q_3 +p) \frac{K_{0\L_0}(q_1)}{q_1^2}
\frac{K_{0\L_0}(q_2)}{q_2^2}\frac{K_{0\L_0}(q_3)}{q_3^2}
\cr &
\times
p_\mu \G_{\mu\mu_1\mu_2\mu_3}^{b\bx}(p;q_1,q_2,q_3)
B_{\mu_1\mu_2\mu_3}(q_1,q_2,q_3;S)\;.
}
\eeq
Assuming for the diagrams with $b$ boxes the asymptotic behaviours
given in Eq.~\re{1boxrel} and using \re{box}-\re{box'} for the box
vertices $B$, we reproduce the assumption for the diagrams with $b+1$ boxes.
In the limit \UV we have then proved the transversality of the photon
for the graphs with the generic structure of Fig.~5.

Finally we have to consider the case of graphs in which more than one photon
is emitted by the fermion loop.
The case with two of such photons ($p$ and $p'$) is obtained by
making the substitution
$
\G^{(1)}_{\mu\mu_1\cdots \mu_n}(p,q_1\cdots q_n) \to
\G^{(1)}_{\mu\nu\mu_1\cdots \mu_n}(p,p',q_1\cdots q_n)\;.
$
in \re{1box}.
In this case the relevant intermediate lines are given only by two photons
and the proof proceeds as before. For more than two external photons emitted by
the first fermion loop we don't have intermediate relevant lines and the proof
reduces to the analysis of the one box contribution and can be done as
in the previous case.

\section {Perturbative renormalization}
Perturbative renormalization is essentially based on power counting.
In four space time dimensions
a given vertex $\G_{c_1 \cdots c_n}$ with $n_A$ photons and $n_\psi$
pairs of fermions has dimension in mass given by
$$
\mbox{dim}\; \G_{c_1 \cdots c_n} = \dim\,.
$$
To prove perturbative renormalizability we shall follow the method of
Ref.~\cite{BDM}. We analyse the UV finiteness of the loop expansion
contributions which are obtained by iterating the integral equations
corresponding to the evolution equations \re{eveq} and the boundary
conditions \re{bc1} and \re{bc2}.

\vskip .3 true cm
\noindent
1. {\it Integral equations}
\vskip .2 true cm

For the relevant couplings we have the integral equations
\bminiG{}
\nome{int1}
\s_i(\L) &=& e\delta_{i,e}+\int_0^\L
\frac {d\l}{\l}
R_i(\l)\,,
\\
\nome{Ri}
R_i(\l) &=& -\half\int_qM_{ba}(q;\l)\, \bG_{ab,i}(-q,q;\l)\,,
\emini
where $\s_e(0)=e$ is the only coupling different from zero.
The auxiliary vertices $\bG_{ab,i}$ are given in Appendix B.

For the irrelevant vertices we have two cases according to whether
$\dim < 0$ or $\dim \ge 0$.
In the first case the vertices satisfy the integral equation
\beq\nome{int2}
\G_{c_1\cdots c_n}(p_1,\cdots p_n;\L)=
-\int_\L^{\L_0} \frac {d\l}{\l}
I_{c_1\cdots c_n}(p_1,\cdots p_n;\l)\,,
\eeq
where
$$
I_{c_1\cdots c_n}(p_1,\cdots p_n;\l)=
-\half\int_qM_{ba}(q;\l)\,
\bG_{ab,c_1 \cdots c_n}(-q,q;p_1,\cdots p_n;\l)\,.
$$
In the second case we have the five vertices $\S_i$ given by
$$
\S_L(p;\L)\,,\;\;
\S_T(p;\L)\,,\;\;
\S_\psi(p;\L)\,, \;\;
\S_\mu(p,p';\L)\,, \;\;
\S_{\mu_1\cdots\mu_4}(p_1,\cdots p_4;\L)\,,
$$
which satisfy the integral equations
\beq\nome{int3}
\S_i(p\cdots ;\L)=
-\int_\L^{\L_0} \frac {d\l}{\l}
I^s_i(p\cdots ;\l) \,,
\eeq
where the subtracted integrands $I^s_i$ are given in Appendix B.
For instance for $\S_L(p;\L)$ we have
\beq\nome{IS2}
\eqalign{
& I^s_L(p;\l)= I_L(p;\l)- I_L(0;\l)-
p^2\, \partial_{p'^2} I_L(p';\l)|_{p'=\bp}
\cr &
=
(p\partial_{p})^4
\int_0^1 \frac{ dx (1-x)^3}{3!x^4} \, I_L(xp;\l)\,
-\half (p\partial_{p'})^2
[I_L(p';\l)|_{p'=\bp}-I_L(p';\l)|_{p'=0}]
\cr&
+2(p\bp)^2(\partial_{\bp^2})^2 I_L(\bp;\l)
\,,
}
\eeq
where
\beq\nome{IS1}
I_L(p;\l)=
-\half\int_qM_{ba}(q;\l)\,
\bG_{ab,L}(-q,q;-p,p;\l)\,.
\eeq
As we see from this expression, since the subtracted vertex vanishes
with its derivative, we can factorize four powers of momentum and
the remaining function has negative dimension.

\vskip .3 true cm
\noindent
2. {\it Perturbative analysis}
\vskip .2 true cm

To prove that the theory is renormalizable we must show
that the integral equations give a finite result in the limit \UV.
This can be done perturbatively by iterating Eq.~\re{int1}, \re{int2}
and \re{int3}, in which we set \UV. In order to see that the integrations
over $\l$ are convergent, we have to estimate the behaviour of the integrands
for large $\l$.
The analysis can be simplified, following Polchinski, by introducing the norm
$$
|f|_\l \equiv \Maxlp |f(p_1,\cdots p_n;\l)|\,,
$$
which allows us to ignore the momentum dependence.
Since the $\L$-dependence is fixed only by the
number of photons and fermions, to simplify the notation, we indicate
in the vertices only the numbers $n_A$ and $n_\psi$
$$
|\G_{c_1\cdots c_n}(p_1\cdots p_4;\L)|_\L \equiv |\G(n_A,n_\psi)|_\L\,.
$$
We shall deduce perturbative renormalization by proving by induction
on the number of loops
that this norm satisfies power counting. Namely, at loop
$\ell$, we assume for large $\L$ the following behaviours
\beq\nome{pwc}
\eqalign{&
\s^{(\ell)}_{m_A}(\L)=\cal{O}(\L^2)\,,\;\;\;\;
\s^{(\ell)}_{m_\psi}(\L)=\cal{O}(\L)\,,
\cr &
\s^{(\ell)}_A(\L)\,,\;\;
\s^{(\ell)}_\psi(\L)\,,\;\;
\s^{(\ell)}_\alpha(\L)\,,\;\;
\s^{(\ell)}_e(\L)\,,\;\;
\s^{(\ell)}_{4}(\L)\,=\,\cal{O}(1)\,,
\cr &
|\G^{(\ell)}(n_A,n_\psi)|_\L\,=\,\cal{O}(\L^{4-n_A-3n_\psi})\,, \;\;\;
\dim <0\,,
\cr &
|\S^{(\ell)}_{T,L}|_\L=\cal{O}(\L^2)\,,\;\;\;\;
|\S^{(\ell)}_{\psi}|_\L=\cal{O}(\L)\,,
\cr &
|\S^{(\ell)}_{\mu}|_\L=\cal{O}(1)\,,\;\;\;\;
|\S^{(\ell)}_{\mu_1\cdots\mu_4}|_\L=\cal{O}(1)\,.
}
\eeq
We neglect for simplicity all possible $\ell$-dependent powers of
$\log\frac{\L}{m}$.
These behaviours are satisfied for the one loop case discussed in section 3.
We now proceed by iteration and prove that the behaviours \re{pwc} are
reproduced at the loop $\ell+1$. To do this we need to give a bound
of the r.h.s. of the integral equation by estimating $R_i(\l)$ in
\re{int1} and the norms of $I_{c_1 \cdots c_n}(p_1,\cdots p_n;\l)$ and
$I^s_i(p, \cdots;\l)$ in \re{int2} and \re{int3}.

First of all we notice that the norm of the auxiliary vertices have
the same behaviours as the corresponding vertices, as can be
seen from their definition. Thus we have
\beq
\eqalign{
&
|\bG^{(\ell)}_{\alpha\beta,c_1 \cdots c_n}(-q,q;p_1,\cdots p_n;\l)|_\l=
\cal{O}(\l^{\dim-3})\,,
\cr &
|\bG^{(\ell)}_{\mu\nu,c_1 \cdots c_n}(-q,q;p_1,\cdots p_n;\l)|_\l=
\cal{O}(\l^{\dim -2})\,.
}
\eeq
Then we observe that the irrelevant vertices with positive dimensions
are given by subtracted auxiliary vertices. These are obtained by
taking derivatives  of the auxiliary vertices with respect to the
external momenta. This reduces the powers of $\l$
\beq\nome{der}
|\partial_{p_i}^m
\G^{(\ell)}_{c_1 \cdots c_n}(p_1,\cdots p_n;\l)|_\l=
\cal{O}(\l^{\dim -m})\,.
\eeq
The proof of this is given for instance in Ref.~\cite{BDM}.
Notice that from the assumptions we find
\beq\nome{neg}
|\Pi^{(\ell)}(q;\l)\frac{1}{\ds{q}+m}|_{\l}=\cal{O}(1)\,,\;\;\;\;
|\Pi^{(\ell)}_{T,L}(q;\l)\frac{1}{q^2}|_{\l}=\cal{O}(1)\,,
\eeq
thus, using \re{Mab'}, $M_{ab}$ gives a factor ${\l}^{-1}$ for
fermions and $\l^{-2}$ for photons. We now obtain from the iterative
solution of the integral equations the norm of vertices and couplings
at loop $\ell+1$.

We discuss first the relevant couplings. Consider for instance the
iterative equation for $\s_{m_A}(\L)$. We have
$$
R^{(\ell)}_{m_A}(\l)=-\half\int_qM^{(\ell')}_{ba}(q;\l)\,
\bG^{(\ell-\ell')}_{abL}(-q,q;0,0;\l)\,,
$$
which gives
$$
R^{(\ell)}_{m_A}(\l)\ltap
 \l^2|\G^{(\ell)}(4,0)|_\l
+\l^3|\G^{(\ell)}(2,1)|_\l
=\cal{O}(\l^2)\,,
$$
where we neglected the self energy contributions due to \re{neg}
and we have used \re{pwc} for the vertex
$|\G^{(\ell)}(n_A,n_\psi)|_\l$. We find then
$$
\s^{(\ell+1)}_{m_A}(\L)\ltap\int_0^\L\frac{d\l}{\l}\;
\left( \l^2|\G^{(\ell)}(4,0)|_\l+\l^3|\G^{(\ell)}(2,1)|_\l
\right) =\cal{O}(\L^2)\,.
$$
Similar results are obtained for the other couplings. Notice that in
this case the integrand grows with $\l$ and the result is dominated by
the upper limit $\L$.

We consider now the case of irrelevant vertices. In this
case the $\l$-integration goes up to infinity (for \UV).
We treat separately the irrelevant vertices with negative and
positive dimensions. For the first ones we have
\beq
\eqalign{ &
|I^{(\ell)}(n_A,n_\psi)|_\l\simeq
\l^2|\G^{(\ell)}(n_A+2,n_\psi)|_\l+\l^3|\G^{(\ell)}(n_A,n_\psi+1)|_\l
=\cal{O}(\l^{4-n_A-3n_\psi})\,,
\cr &
|\G^{(\ell+1)}(n_A,n_\psi)|_\L\ltap
\int_\L^\infty\frac{d\l}{\l}|I^{(\ell)}(n_A,n_\psi)|_\l
=\cal{O}(\L^{4-n_A-3n_\psi})\,.
}
\eeq
Again self energy contributions can be ignored. Since
$4-n_A-3n_\psi<0$ the integration over $\l$ is convergent and the integral
is dominated by the lower limit, thus reproducing
at loop $\ell+1$ the Ansatz \re{pwc}.

We now study the irrelevant vertices with non negative dimension.
Consider for instance the case of $\S^{(\ell+1)}_L(p;\L)$ (see \re{IS1}
and \re{IS2}).
Due to the subtractions the integrand can be expressed as fourth
derivative with respect to the external momenta for which we can use \re{der}.
We have then
\beq\nome{tay}
|I^{s(\ell)}_L(p;\l)|_\l\sim p^4|\partial^4 I^{(\ell)}_L(p;\l)|_\l
=p^4\cal{O}(\l^{-2})
\eeq
for $p^2<\l^2$. We find
$$
|\S^{(\ell+1)}_L(p;\L)|_\L
\le\L^4\int_\L^\infty\frac{d\l}{\l}
|\partial^4 I^{(\ell)}_L(p;\l)|_\l
=\cal{O}(\L^{2})\,.
$$
We see here that the subtractions make the $\l$-integration convergent
and dominated by the lower limit $\L$. The positive power of $\L$ in
the Ansatz \re{pwc} is recovered from $\L^4$ coming by maximizing
the factor $p^4$ in \re{tay}.

\section{Conclusions}
In this paper we have analyzed how the Ward identities emerge in the
Wilson renormalization group formulation of QED.
The main feature is that this gauge symmetry must be implemented at
the level of the relevant part of the physical effective action.
This implies that the seven relevant couplings $\s_i(\L)$ must be fixed
at $\L=0$, according to \re{action} or \re{bc1}.
We have shown that in perturbation theory the renormalization group
equations ensure that the full effective action at $\L=0$ and \UV
satisfies the Ward identities.
This fact is deeply connected to the renormalizability of the theory,
\ie to the large momentum behaviour of the integrands of Feynman
diagrams.
The UV cutoff $\L_0$ must be kept when we study
graphs which do not include the subtractions imposed by the boundary
conditions in \re{action}.
We find that the Ward identity violating contributions involve differences
of the cutoff functions at $\L=0$ which force the
integration momenta around the UV cutoff.
Thus their behaviours for large $\L_0$ are obtained by
estimating the large momentum limit of Feynman diagram integrands,
which are given by the Weinberg theorem.
For the vertex identities, the result is that for \UV
the violation,
${\cal W}^{\cal G}(p,p';\L_0)$, is vanishing as an inverse power
of the cutoff.
This is the reason why we can avoid a detailed study of
other subtractions, which behave logarithmically in $\L_0$, and the
relative forest expansion.
For the photon transversality, we have that for \UV the possible non
vanishing longitudinal contributions are given by relevant contributions,
\ie by polynomial at most of quadratic degree in the external momenta, which
are automatically cancelled by subtractions.

We want to emphasize that, in order to satisfy the Ward identities,
one must require not only \UV but also $\L=0$.
For $\L \ne 0$ the difference of cutoff functions does not force the
integration momentum in the UV region only and the violation of the Ward
identities would not be polynomial in the external momenta.
One can verify at the one loop level that even for \UV the
Ward identities are not satisfied as long as $\L \ne 0$.
See for instance Eq.~\re{WIvv} for the vertex identity.

In conclusion the functional $\G[A,\psi, \bar \psi;\L]$ does not
satisfy Ward identities except at the physical point $\L=0$ and \UV.
At the UV scale $\L=\L_0$ this functional is given only by the
relevant part, \ie by the seven couplings $\s_i(\L_0)$ which are
the couplings $\s^B_i$ in the effective
Lagrangian at the UV scale in Eq.~\re{intac}.
Although these couplings do not satisfy Ward identities, they are
functionally related since they lead to the physical effective action
which fulfils these identities.
In Eqs.~\re{bare1},\re{bare2},\re{bare3} and \re{bare4} we have
computed the seven couplings $\s_i(\L)$ at one loop for large $\L$,
using a sharp momentum cutoff.

In the last part of the paper we have given  a simple proof of
perturbative renormalizability of QED in this framework.
We used the same method of Ref.~\cite{BDM}.
The proof is based on the fact that Feynman graphs obtained by iteratively
solving the evolution equations are organized in such a way
that the loop momenta are ordered.
It is then possible to analyze their ultraviolet
behaviour by iterative methods. The necessary
subtractions and the corresponding counterterms are automatically
generated in the process of fixing the physical conditions for
the ``relevant'' vertices. The proof of perturbative renormalizability
is simply based on dimensional arguments and does not require the
usual analysis of topological properties of Feynman graphs.

\vspace{3mm}\noindent{\large\bf Acknowledgements}

We have benefited greatly from discussions with C. Becchi, G. Bottazzi
and M. Tonin.

\newpage

{\bf Appendix A}
\vskip 1 true cm

We derive here the evolution equation for the cutoff effective action.
Taking into account that
$$
\dL \G[\Phi;\L]=-\dL W[J;\L]+\dL W[0;\L] \,,
$$
the evolution equation for the cutoff effective action is given by
\beq\nome{app1}
\eqalign{
\dL \biggr\{ &\G[\Phi;\L]-\half \int_p
\Phi_a(-p) D^{-1}_{ab}(p;\L)\,\Phi_b(p)\biggr\}
\cr&
=\dL W[0;\L]
\;+\;
\half (2\pi)^8 \int_q \DL{ D^{-1}_{ba}(q;\L)} \,
\frac { \delta^2 W[J;\L]}{\delta J_b(q)\delta J_a(-q)} \,
\;.
}
\eeq
The functional in the integrand is the inverse of
$\delta^2\G[\Phi;\L]/\delta\Phi_a(q)\delta\Phi_b(q')$
\beq\nome{inv}
\delta^4(q+q')\, \delta_{ab}=(2\pi)^{8}
\int_{q''}
\;
\frac{\delta^2 W[J]}{\delta J_b(q')\delta J_c(-q'')}
\;
\frac{\delta^2 \Gamma[\Phi]}{\delta\Phi_c(q'')\delta\Phi_a(q)}
\;
(-)^{\delta_a}
\;,
\eeq
where $\delta_a$ is one if $a$ is a fermion index and zero otherwise.
We isolate, similarly to Eq.~\re{inv2}, the two point function
contribution in the functional $W[J;\L]$
$$
(2\pi)^8\frac {\delta^2\, W[J;\L]}{\delta J_a(q)\delta J_b(q')}=
(2\pi)^4\delta^4(q+q')\, (-)^{\delta_a} \, \Delta_{ba}(q;\L)+
W_{ba}^{int}[q',q;J]\,.
$$
Notice that the two point function contribution
cancels the constant term $\L \partial W[0;\L]/\partial \L$ in
\re{app1}.
Then, from \re{inv} we express $W_{ab}^{int}$ as functional of $\Phi(p)$
$$
W_{ab} ^{int}[-q,q';J]=-
\Delta_{ac} (q;\L) \, \bG_{cd} [-q,q';\Phi]\,\Delta_{db}(q';\L)
\,,
$$
where the auxiliary functional $\bG$ satisfies the Eq.~\re{bG}
(see Fig.~1).
In conclusion we obtain Eq.~\re{eveq}.

\newpage

{\bf Appendix B}
\vskip 1 true cm

The auxiliary vertices $\bG_{ab,i}$ for the equation \re{Ri} are given by
\beq
\eqalign{
&
\bG_{ab,m_\psi}(-q,q;\l)=
-i\frac{1}{4}\delta_{\alpha\beta}
\bG_{ab,\alpha\beta}(-q,q;-\bp,\bp;\l)
\;\;\;\;\;\ds{\bp}=-m
\,,
\cr &
\bG_{ab,\psi}(-q,q;\l)=
i\frac{1}{16}(\g_\mu)_{\beta\alpha}\frac{\partial}{\partial p'_\mu}
\bG_{ab,\alpha\beta}(-q,q;-p',p';\l)|_{p'=0}
\,,
\cr &
\bG_{ab,e}(-q,q;\l)=
i\frac{1}{16}(\g_\mu)_{\beta\alpha}
\bG_{ab,\alpha\beta\mu}(-q,q;0,0,0;\l)
\,,
\cr &
\bG_{ab,m_A}(-q,q;\l)=
\bG_{ab,L}(-q,q;0,0;\l)
\,,
\cr &
\bG_{ab,\alpha}(-q,q;\l)=
\frac{\partial}{\partial p^2}
\bG_{ab,L}(-q,q;-p,p;\l)
|_{p=\bp}\;\;\;\;\;\bp^2=\mu^2
\,,
\cr &
\bG_{ab,A}(-q,q;\l)=
\frac{\partial}{\partial p^2}
\bG_{ab,T}(-q,q;-p,p;\l)
|_{p=\bp}\;\;\;\;\;\bp^2=\mu^2
\,,
\cr &
\bG_{ab,4}(-q,q;\l)=
\frac{1}{24}
\bG_{ab,\mu\mu\nu\nu}(-q,q;0,0,0,0;\l)
\,,
}
\eeq
where we introduced the notations
$$
\bG_{ab,L}=
\frac{p_\mu p_\nu}{p^2}
\bG_{ab,\mu\nu}\,,
\;\;\;\;\;\;\;\;\;\;
\bG_{ab,T}=
\frac{1}{3}(\delta_{\mu\nu}-4\frac{p_\mu p_\nu}{p^2})
\bG_{ab,\mu\nu}\,.
$$
The integrands $I_i^s$ for the five vertices $\S_i$ (see \re{int3})
are given by
$$
I^s_i(p_1\cdots p_n;\l)=
-\half\int_q M_{ba}(q;\l) \bG^s_{ab,i}(-q,q;p_1\cdots p_n;\l)\,,
$$
where the subtracted auxiliary vertices are
\beq
\eqalign{
&
\bG^s_{ab,\psi}=
\bG_{ab,\alpha\beta}(-q,q;-p,p;\l)
-\bG_{ab,\alpha\beta}(-q,q;-\bp,\bp;\l)
\cr & \;\;\;\;\;\;\;
-(p_\mu-\frac{m}{4}\g_\mu)_{\alpha\alpha'}\frac{\partial}{\partial p'}
\bG_{ab,\alpha'\beta}(-q,q;-p',p';\l)|_{p'=0}
\;
\sim\, (p\partial)^2\,\bG
\;\;\;\;\;\;\ds{\bp}=-m
\,,
\cr &
\bG^s_{ab,\mu}=
\bG_{ab,\alpha\beta\mu}(-q,q;-p,p',p-p';\l)
-\bG_{ab,\alpha\beta\mu}(-q,q;0,0,0;\l)\;
\sim\, p\partial\,\bG
\,,
\cr &
\bG^s_{ab,L}=
\bG_{ab,L}(-q,q;-p,p;\l)
-\bG_{ab,L}(-q,q;0,0;\l)
\cr & \;\;\;\;\;\;\;
-p^2\frac{\partial}{\partial p'^2}
\bG_{ab,L}(-q,q;-p',p';\l)|_{p'^2=\mu^2}
\;
\sim\, (p\partial)^4\,\bG
\,,
\cr &
\bG^s_{ab,T}=
\bG_{ab,T}(-q,q;-p,p;\l)
-p^2\frac{\partial}{\partial p'^2}
\bG_{ab,T}(-q,q;-p',p';\l)|_{p'^2=\mu^2}
\;\;
\sim\, (p\partial)^4\,\bG
\,,
\cr &
\bG^s_{ab,\mu_1\cdots\mu_4}=
\bG_{ab,\mu_1\cdots\mu_4}(-q,q;p_1\cdots p_4;\l)
-\bG_{ab,\mu_1\cdots\mu_4}(-q,q;0\cdots 0;\l)
\;
\sim\, (p\partial)^2\,\bG
\,.
}
\eeq

\eject

\eject
\newpage
\begin{figcap}

\item
Graphical representation of Eq.~\re{bG} defining the auxiliary
functional $\bG_{ab}[q,q';\Phi]$. Solid lines represent in this case
both fermions and photons. The dashed blob  represents the functional
$\bG_{ab}^{int}[q,q';\Phi]$ defined in Eq.~\re{inv2}.
\item
a) Graphical representation of the auxiliary vertex at zero loop for
the electron emitting $n$ photons.\\
b) Auxiliary vertex giving the electron propagator  at one loop (see
Eq.\re{S1}).
\item
a) General graph for the fermion propagator. The function $G$ corresponds
to the sum of connected and disconnected graphs.\\
b) Example in which the function $G$ is disconnected. No fermion self energy
subgraphs are here included.\\
c) Example in which the function $G$ is disconnected. In this case $G'_c$
gives subgraphs of fermion self energy corrections.
\item
General graph for the fermion-photon vertex.
\item
General graph for a photon $p$ emitted by a fermion loop.
The other external lines in $S$ are not emitted by this loop.
\item
Expansion for the graph of Fig.~5. The  box is defined as the sum
of diagrams for which a vertical line cuts intermediate propagators
which are not ``relevant lines'', \ie a single fermion pair or three
photons only.

\end{figcap}

\end{document}